\documentstyle[11pt]{article}
\begin{document}

\begin{titlepage}
\begin{center}
{\LARGE\bf MADCAP --- The Microwave Anisotropy Dataset Computational Analysis Package}\\
\vspace{.3in}
{\Large Julian Borrill} \\
\vspace{.15in}
{\em Scientific Computing Group, \\
National Energy Research Scientific Computing Center, \\
Lawrence Berkeley National Laboratory \\
\& \\
Center for Particle Astrophysics, \\
University of California at Berkeley
}
\end{center}
\baselineskip=24pt
\begin{abstract}

In the standard model of cosmology the universe starts with a hot Big
Bang. As the universe expands it cools, and after 300,000 years it
drops below the ionisation temperature of hydrogen. The previously
free electrons become bound to protons, and with no electrons for the
photons to scatter off they continue undeflected to us today. This
image of the surface of last-scattering is what we call the Cosmic
(because it fills the universe) Microwave (because of the frequency at
which its black body spectrum peaks today) Background (because it
originates behind all other light sources). Despite its stunning
uniformity - isotropic to a few parts in a million - it is the tiny
perturbations in the CMB that give us an unprecedented view of the
early universe. First detected by the COBE satellite in 1991, these
anisotropies are an imprint of the primordial density fluctuations
needed to seed the development of gravitationally bound objects in the
universe, and are potentially the most powerful discriminant between
cosmological models.

Realizing the extraordinary scientific potential of the CMB requires
precise measurements of these tiny anisotropies over a significant
fraction of the sky at very high resolution. The analysis of the
resulting datasets is a serious computational challenge. Existing
algorithms require terabytes of memory and hundreds of years of CPU
time. We must therefore both maximize our resources by moving to
supercomputers and minimize our requirements by algorithmic
development. Here we will outline the nature of the challenge, present
our current optimal algorithm, discuss its implementation - as the
MADCAP software package - and its application to data from the North
American test flight of the joint Italian-U.S. BOOMERanG experiment on
the Cray T3E at NERSC and CINECA.

A documented $\beta$-release of MADCAP is publicly available at
\begin{center}
http://cfpa.berkeley.edu/$\sim$borrill/cmb/madcap.html
\end{center}

\end{abstract}
\end{titlepage}
 
\section{Introduction}

The current standard model of cosmology starts with a hot Big
Bang. Whilst there is still debate about what this actually means, it
is generally agreed is that what emerges is a very hot, expanding,
space-time --- the Universe. As the Universe expands its temperature
falls and about 300,000 years after the Big Bang it cools to below the
ionisation temperature of hydrogen and the previously free electrons
become bound to protons. With no electrons to scatter off, the photons
propagate undeflected through space to the present. When we observe
this radiation today we are seeing the universe as it was when it was
1/40,000th of its present age. This image of the epoch when the
primordial photons last scattered is what we call the Cosmic Microwave
Background (CMB) radiation.

Detected serendipitously in 1969, the CMB was considered the decisive
argument in the debate of the day between the `Steady State' and `Big
Bang' cosmologies. Today, extraordinary instruments are measuring the
tiniest variations in the CMB photons' temperature in different
directions, and the results they are giving hold the promise of
settling this generation's cosmological debates
\cite{PARAMETERS}. They have the potential to describe the geometry of
space-time, showing whether straight lines continue forever or not;
they can tell us how much mass and energy the universe contains, and
the forms that it takes; and they may shed light the kinds of things
that could have happened in the very first moments of the Big Bang.

As observers began to search for fluctuations in the CMB temperature
they soon discovered that it was astonishingly uniform -- 2.735 Kelvin
in every direction. Building ever more sensitive detectors, the first
variations weren't discovered until the milliKelvin regime. However
these were due our galaxy's motion through space, a Doppler effect
making the universe appear hotter in the direction in which we are
going and colder in the direction from which we have come. Finally in
1992 the COBE satellite team reported intrinsic spatial variation in
the CMB's temperature of around $\pm 30 \mu K$ when averaged over
patches of sky approximately $10^{\rm o}$ across.

Since that detection the focus has been on measuring the extent of the
variation with ever greater precision on ever smaller angular
scales. Both the absolute and the relative power of the fluctuations
on different angular scales contain a wealth of cosmological
information. For example, we expect to see a peak in the power at the
angular scale corresponding to the horizon size (the limiting scale
for coherent physical processes) at last scattering. The apparent
angular size on the sky today of a particular physical scale in the
past depends on the geometry of the Universe -- decreasing in size as
the curvature of space increases. A measurement of the location of the
peak in the angular power spectrum therefore tells us whether the
Universe is open, flat or closed. Similarly the height of this peak is
related to the total energy density in the universe, and how it is
distributed. The presence or absence of lower secondary peaks, at
resonances of the fundamental scale, may allow us to rule out some
classes of theories of the origins of the first density perturbations
in the universe.

Making an observation with small enough statistical and systematic
error bars to resolve the detailed shape of this angular power
spectrum requires very sensitive detectors scanning a significant
fraction of the sky at very high resolution. These conditions are now
being met for the first time in balloon-borne experiments such the
joint Italian--U.S. BOOMERanG project. Lifted into the stratosphere to
minimize atmospheric interference, this experiment measures changes in
the voltage across an extremely sensitive bolometer, cooled to a few
milliKelvin, as it scans the sky. Since the voltage across a bolometer
depends on it's temperature, hidden within such a data set is a
measurement of CMB temperature fluctuations.

Reducing tens of millions of individual observations to an angular
power spectrum of a thousand multipoles is a computationally
challenging task. Each observation contains detector noise as well as
signal on the sky; the signal on the sky includes not only the CMB but
also foreground sources of microwave radiation such as interstellar
dust; and both the noise and the signal components of nearby
observations are correlated. At present, except in very restrictive
circumstances, algorithms to solve the equations relating the
time-ordered data to a map of the sky temperature, and then to the
angular power spectrum, scale as the number of map pixels squared in
memory and cubed in floating point operation count. The COBE map
contained only a few thousand pixels, but current balloon observations
are generating maps with tens and hundreds of thousands pixels, and
the MAP and PLANCK satellites --- to be launched in 2000 and 2007
respectively --- will increase this to millions and tens of millions.

To realize the full potential of these CMB observations we
simultaneously need to maximize our computational resources (by moving
to supercomputers) and to minimize our computational requirements (by
optimizing our algorithms and implementations). Here we present our
current optimal algorithm and discuss its implementation in the MADCAP
software package and describe its application at NERSC and CINECA to
data from the North American test flight of BOOMERanG. For simplicity
we will consider an observation from a single detector with no
significant foreground contamination --- a situation that was realized
in practice in the analysis of the best channel of the BOOMERanG North
America data.

In what follows vectors and matrices are written in plain fonts in the
time domain and italic fonts in the pixel domain.

\section{From The Time-Ordered Data To The Map}

\subsection{Algorithm}

Our first step is to translate the observation from the temporal to
the spatial domain --- to make a map \cite{MAP}. Knowing where the
detector was pointing, $(\theta_{t}, \psi_{t})$, at each of the ${\cal
N}_{t}$ observation, and dividing the sky into ${\cal N}_{p}$ pixels,
we can construct an ${\cal N}_{t} \times {\cal N}_{p}$ pointing matrix
${\rm A}$ whose entries give the weight of pixel $p$ in observation
$t$. For a total power scanning experiment such as BOOMERanG this has
a particularly simple form
\begin{equation}
{\rm A}_{t p} = \left\{ \begin{array}{ll}
	        1 & {\rm if} \;\;\; (\theta_{t}, \psi_{t}) \in p \\
                0 & {\rm otherwise}
	                 \end{array}
                \right.
\end{equation}
while a more complex observing strategy would give a correspondingly
complex structure.

The time-ordered data vector can now be written
\begin{equation}
\label{eDSN}
{\rm d} = {\rm A} \, s + {\rm n}
\end{equation}
in terms of the pixelised CMB signal $s$ and time-stream noise
${\rm n}$. Under the assumption of Gaussianity, the noise probability
distribution is
\begin{equation}
\label{eTTNPD}
{\rm P}({\rm n}) = (2 \pi)^{-{\cal N}_{t}/2} \exp \left\{ -\frac{1}{2} \left(
{\rm n}^{\rm T} \, {\rm N}^{-1} \, {\rm n} + {\rm Tr} \left[ \ln {\rm N} \right] \right) \right\}
\end{equation}
where ${\rm N}$ is the time-time noise correlation matrix given by
\begin{equation}
{\rm N} \equiv \langle {\rm n} \, {\rm n}^{\rm T} \rangle
\end{equation}
We can now use equation (\ref{eDSN}) to substitute for the noise in
equation (\ref{eTTNPD}), so the probability that, with a particular
underlying CMB signal, we would have obtained the observed
time-ordered data is
\begin{equation}
{\rm P}({\rm d} | s) = (2 \pi)^{-{\cal N}_{t}/2} \exp \left\{ -\frac{1}{2} \left(
({\rm d} - {\rm A} \, s)^{\rm T} \, {\rm N}^{-1} \, ({\rm d} - {\rm A} \, s) + {\rm Tr} \left[ \ln {\rm N} \right] \right)
\right\}
\end{equation}
Assuming that all CMB signals are {\em a priori} equally likely, this
is proportional to the likelihood of the CMB signal given the
time-ordered data.  Maximizing over $s$ now gives the maximum
likelihood pixelized data (or map) $d$
\begin{equation}
\label{eMAP}
d = \left( {\rm A}^{\rm T} \, {\rm N}^{-1} \, {\rm A} \right)^{-1} {\rm A}^{\rm T} \, {\rm N}^{-1} \, {\rm d}
\end{equation}

Substituting back for the time-ordered data in equation (\ref{eMAP})
we recover the obvious fact that this pixelized data is the sum of the
true CMB signal and some residual pixelized noise
\begin{eqnarray}
d & = & \left( {\rm A}^{\rm T} \, {\rm N}^{-1} \, {\rm A} \right)^{-1} \, {\rm A}^{\rm T} \, {\rm N}^{-1} \, ({\rm A} \, s + {\rm n}) \nonumber \\
        & = & s + n
\end{eqnarray}
where this pixel noise
\begin{equation}
n = \left( {\rm A}^{\rm T} \, {\rm N}^{-1} \, {\rm A} \right)^{-1} \, {\rm A}^{\rm T} \, {\rm N}^{-1} \, {\rm n}
\end{equation}
has correlations given by
\begin{eqnarray}
N & = & \langle n \, n^{\rm T} \rangle \nonumber \\
  & = & \left( {\rm A}^{\rm T} \, {\rm N}^{-1} \, {\rm A} \right)^{-1}
\end{eqnarray}
The map-making algorithm can therefore be divided into two steps,
\begin{enumerate}
\item[M1 --] construct the inverse pixel-pixel noise correlation matrix and noise-weighted map
$$ N^{-1} = {\rm A}^{\rm T} \, {\rm N}^{-1} \, {\rm A} $$
$$ z \equiv N^{-1} \, d = {\rm A}^{\rm T} \, {\rm N}^{-1} \, {\rm d} $$
\item[M2 --] solve for the pixelized data
$$ d = (N^{-1})^{-1} \, z $$
\end{enumerate}
which are encoded in the two map-making modules --- {\em
inv\_pp\_noise.c} and {\em p\_data.c} --- in MADCAP.

\subsection{Implementation}

The first half of Table 1 shows the computational cost of a brute
force implementation of each of the steps in the map-making algorithm
(recall that multiplying an $[a \times b]$ matrix and a $[b \times c]$
matrix in general requires $2 \, a \, b \, c$ operations). For the
BOOMERanG North America data --- with ${\cal N}_{t} \sim 1.5 \times
10^{6}$ and ${\cal N}_{p} \sim 2.4 \times 10^{4}$ --- simply making
the map would require 9 Tb disc space (storing data in 4-byte
precision), 18 Tb RAM\footnote{Since all algorithms here will be
operation-count limited we always assume in-core
implementations. Out-of-core methods would reduce memory requirements
but prohibitively increase the run-time overhead, for example in the
cache hit cost of moving from level 3 (matrix-matrix) to level 1
(vector-vector) BLAS.}  (doing all calculations in 8-byte precision)
and $10^{17}$ floating point operations.

\begin{table*}[ht!]
\centering
\begin{tabular}{ccccccc}
Calculation & \multicolumn{3}{c}{Brute Force} & \multicolumn{3}{c}{Structure-Exploiting} \\
& Disc & RAM & Flops & Disc & RAM & Flops \\
\hline
\vspace*{-0.1in}&&&&&\\
M1 &
$4 \, {\cal N}_{t}^{2}$ & 
$8 \, {\cal N}_{t}^{2}$ & 
$2 \, {\cal N}_{t}^{2} \, {\cal N}_{p}$ & 
$4 \, ({\cal N}_{p}^{2} + {\cal N}_{t})$ &
$8 \, ({\cal N}_{p}^{2} + {\cal N}_{t})$ & 
$3 {\cal N}_{\tau} \, {\cal N}_{t}$ \\
M2 &
$4 \, {\cal N}_{p}^{2}$ & 
$8 \, {\cal N}_{p}^{2}$ & 
$(2 + \frac{2}{3}) \, {\cal N}_{p}^{3}$ & 
$4 \, {\cal N}_{p}^{2}$ & 
$8 \, {\cal N}_{p}^{2}$ & 
$(2 + \frac{2}{3}) \, {\cal N}_{p}^{3}$ \\
\vspace*{-0.1in}&&&&&\\
\hline
\vspace*{-0.1in}&&&&&\\
Total &
$4 \, {\cal N}_{t}^{2}$ & 
$8 \, {\cal N}_{t}^{2}$ & 
$2 \, {\cal N}_{t}^{2} \, {\cal N}_{p}$ & 
$4 \, ({\cal N}_{p}^{2} + {\cal N}_{t})$ & 
$8 \, ({\cal N}_{p}^{2} + {\cal N}_{t})$ & 
$(2 + \frac{2}{3}) \, {\cal N}_{p}^{3}$ \\
\end{tabular}
\caption{Computational requirements for the map-making algorithm}
\end{table*}

Fortunately there are two crucial structural features to be exploited
here. As noted above, for a simple scanning experiment like BOOMERanG
the pointing matrix $A$ is very sparse, with only a single 1 in each
row. Moreover, the inverse time-time noise correlations are (by fiat)
both stationary and fall to zero beyond some time-separation much
shorter than the duration of the observation
\begin{eqnarray}
N^{-1}_{t t'} & = & f(|t - t'|) \nonumber \\
              & = & 0 \;\;\;\;\;\; \forall \;\;\; |t - t'| > \tau \ll {\cal N}_{t}
\end{eqnarray}
so that the inverse time-time noise correlation matrix is symmetric
and band-diagonal, with bandwidth ${\cal N}_{\tau} = 2 \tau + 1$. The
second half of Table 1 shows the impact of exploiting this structure
on the cost of each step. The limiting step is now no longer
constructing the inverse pixel-pixel noise correlation matrix but
inverting it and solving for the map. Although this inversion is not
necessary to obtain just the map, the current power spectrum algorithm
requires the pixel-pixel noise correlation matrix itself and not simply
its inverse. We therefore Cholesky decompose the positive definite
matrix $N^{-1}$, and use the decomposition both to solve for the map
and to calculate $N$. For the same dataset making the map now takes
2.3 Gb of disc, 4.6 Gb of RAM, and $3.7 \times 10^{13}$ flops.

Although the final analysis of the data is dominated by the second
step, `quick and dirty' systematics tests can be performed at much
lower map resolution (ie. with much larger pixels) to reduce ${\cal
N}_{p}$ by up to an order of magnitude. At this point it becomes
important to optimize the first step too. The structure exploiting
algorithm reduces these calculations to:\\
\hspace*{0.25in} for each observation (at time $t$, of pixel $p$) \\
\hspace*{0.50in} for each observation within $\tau$ of it (at time $t'$, of pixel $p'$) \\
\hspace*{0.75in} add $f(|t - t'|)$ to $N^{-1}_{p p'}$ \\
\hspace*{0.75in} add $f(|t - t'|) \, d_{t'}$ to $z_{p}$ \\ 
Dividing this into blocks of contiguous $p$-pixels allows each
processor to work independently. However the number of times a pixel
is observed can vary from a few to thousands, so a simple division
into equal numbers of $p$-pixels on each processor can be very poorly
load-balanced. The MADCAP implementation therefore starts by
determining the number of operations required for each $p$-pixel and
dividing them among the processors in blocks which are still
contiguous but whose numbers of $p$-pixels vary so that the total
operation count in each block is as close to the mean count per
processor as possible. Although this means that the memory requirement
per processor varies (and is not known exactly in advance) for the
BOOMERanG North America data it reduced the run-time of these steps by
a factor of two or more.

\section{From The Map To The Power Spectrum}

\subsection{Algorithm}

We now want to move to a basis where the CMB observation can be
compared with the predictions of various cosmological theories --- the
angular power spectrum. We decompose the CMB signal at each pixel in
spherical harmonics
\begin{equation}
s_{p} = \sum_{l m} \, a_{l m} \, B_{l} \, Y_{l m}(\theta_p, \psi_p)
\end{equation}
where $B$ is the pattern of the observation beam (assumed to be
circularly symmetric) in $l$-space. The correlations between such
signals then become
\begin{equation}
S_{p p'} \equiv \langle s_{p} \, s_{p'} \rangle = \sum_{l m} \, \sum_{l' m'} \,
\langle a_{l m} \, a_{l' m'} \rangle \, B_{l} \, B_{l'} \, Y_{l m}(\theta_p, \psi_p) \, Y_{l'
m'}(\theta_{p'}, \psi_{p'})
\end{equation}
For isotropic fluctuations the correlations depend only on the angular
separation
\begin{equation}
\langle a_{l m} \, a_{l' m'} \rangle = C_{l} \, \delta_{l l'} \, \delta_{m m'}
\end{equation}
and the pixel-pixel signal correlation matrix entries become
\begin{equation}
S_{p p'} = \sum_{l} \, \frac{2 l + 1}{4 \pi} \, B_{l}^{2} \, C_{l} \, P_{l}(\chi_{p p'})
\end{equation}
where $P_{l}$ is the Legendre polynomial and $\chi_{p p'}$ the angle
between the pixel pair $p, p'$. These $C_{l}$ multipole powers
completely characterize a Gaussian CMB, and are an otherwise
model-independent basis in which to compare theory with
observations. The finite beam size means that any experiment has a
maximum multipole sensitivity, above which all power is
beam-smeared. Coupled with incomplete sky coverage, this means that in
practice the $C_{l}$ that we extract do {\em not} form a complete
orthonormal basis. We therefore group the ${\cal N}_{l}$ accessible
multipoles into ${\cal N}_{b}$ bins, adopting a fixed spectral shape
function $C_{l}^{s}$ and characterizing the CMB signal by its bin
powers $C_{b}$ with
\begin{equation}
C_{l} = C_{b: l \in b} \, C_{l}^{s}
\end{equation}

Since the signal and noise are assumed to be realizations of
independent Gaussian processes the pixel-pixel map correlations are
\begin{eqnarray}
D & \equiv & \langle d \, d^{\rm T} \rangle \nonumber \\
  & = & \langle s \, s^{\rm T} \rangle + \langle n \, n^{\rm T} \rangle \nonumber \\
  & = & S + N
\end{eqnarray}
and the probability distribution of the map given a particular power
spectrum $C$ is now
\begin{equation}
\label{eMLA}
P(d | C) = (2 \pi)^{-{\cal N}_{p}/2} \exp \left\{ -\frac{1}{2} \left(
d^{\rm T} \, D^{-1} \, d + {\rm Tr} \left[ \ln D \right] \right)
\right\}
\end{equation}
Assuming a uniform prior for the spectra, this is proportional to the
likelihood of the power spectrum given the map. Maximizing this over
$C$ then gives us the required result, namely the most likely
CMB power spectrum underlying the original observation ${\rm d}$.

Finding the maximum of the likelihood function of equation
(\ref{eMLA}) is a much harder problem than making the map. Since there
is no closed-form solution corresponding to equation (\ref{eMAP}) we
must find both a fast way to evaluate the likelihood function at a
point, and an efficient way to search the ${\cal N}_{b}$-dimensional
parameter space for the peak. The fastest general method extant is to
use Newton-Raphson iteration to find the zero of the derivative of the
logarithm of the likelihood function \cite{MLA}. If the log likelihood
function
\begin{equation}
\label{eLL}
{\cal L}(C) = - \frac{1}{2} \left( d^{\rm T} \, D^{-1} \, d + {\rm Tr} \left[ \ln D \right] \right)
\end{equation}
were quadratic, then starting from some initial guess at the maximum
likelihood power spectrum $C_{o}$ the correction $\delta {\cal
C}_{o}$ that would take us to the true peak would simply be
\begin{equation}
\label{eDC}
\delta C_{o} = - \left( \left[ \frac{\partial^2 {\cal L}}{\partial C^{2}} \right]^{-1} 
\frac{\partial {\cal L}}{\partial C} \right)_{C = C_{o}}
\end{equation}
Since the log likelihood function is not quadratic, we now take
\begin{equation}
C_{1} = C_{o} + \delta C_{o}
\end{equation}
and iterate until $\delta C_{n} \sim 0$ to the desired
accuracy. Because any function is approximately quadratic near a peak,
if we start searching sufficiently close to a peak this algorithm will
converge to it. Of course there is no guarantee that it will be the
global maximum, and in general there is no certainty about what
`sufficiently close' means in practice. However experience to date
suggests that the log likelihood function is sufficiently strongly
singly peaked to allow us to use this algorithm with some confidence.

The core of the algorithm is then to calculate the first two
derivatives of the log likelihood function with respect to the
multipole bin powers
\begin{eqnarray}
\frac{\partial {\cal L}}{\partial C_{b}} & = & 
\frac{1}{2} \left( d^{\rm T} \, D^{-1} \, \frac{\partial S}{\partial C_{b}} \, D^{-1} \, d
- {\rm Tr} \left[ D^{-1} \, \frac{\partial S}{\partial C_{b}} \right] \right) \\ 
\frac{\partial^2 {\cal L}}{\partial C_{b} \partial C_{b'}} & = & 
- d^{\rm T} \, D^{-1} \, \frac{\partial S}{\partial C_{b}} \, D^{-1} \, 
\frac{\partial S}{\partial C_{b'}} \, D^{-1} \, d +
\frac{1}{2} {\rm Tr} \left[ D^{-1} \, \frac{\partial S}{\partial C_{b}} \, D^{-1} \,
\frac{\partial S}{\partial C_{b'}}\right]
\end{eqnarray}

\newpage

Each iteration of the power-spectrum extraction algorithm can
therefore be divided into six steps,
\begin{enumerate}
\item[P1 --] Calculate the ${\cal N}_{b}$ pixel-pixel signal correlation bin derivative
matrices
$$\frac{\partial S}{\partial C_{b}} = \sum_{l \in b} \, \frac{2 l + 1}{4 \pi} \, B_{l}^{2} \, C_{l}^{s} \, P_{l}(\chi_{p p'})$$
\item[P2 --] Construct the pixel-pixel map correlation matrix, Cholesky decompose it, and triangular solve for the data-weighted map
$$D = N + \sum_{b} C_{b} \frac{\partial S}{\partial C_{b}} = L \, L^{\rm T} \;\;\;\; \& \;\;\;\;
L \, L^{\rm T} \, z = d$$
\item[P3 --] Triangular solve the linear systems
$$L \, L^{\rm T} \, W_{b} = \frac{\partial S}{\partial C_{b}}$$
\item[P4 --] Assemble the first derivative
$$\frac{\partial {\cal L}}{\partial C_{b}} = \frac{1}{2} \left( d^{\rm T} \, W_{b} \, z - {\rm Tr} \left[ W_{b} \right] \right)$$
\item[P5 --] Assemble the second derivative
$$\frac{\partial^2 {\cal L}}{\partial C_{b} \partial C_{b'}} = - d^{\rm T} \, W_{b} \, W_{b'} \, z + \frac{1}{2} {\rm Tr} \left[ W_{b} \, W_{b'} \right]$$
\item[P6 --] Calculate the spectral correction
$$\delta C = - \left[ \frac{\partial^2 {\cal L}}{\partial C^{2}} \right]^{-1} 
\frac{\partial {\cal L}}{\partial C} $$
\end{enumerate}
which are encoded in the six power-spectrum extraction modules ---
{\em pp\_signal.c}, {\em L.c}, {\em trisolve.c}, {\em dLdC.c}, {\em
d2LdC2.c} and {\em dC.c} --- in MADCAP.

\subsection{Implementation}

Table 2 shows the computational cost of each of the steps in each
iteration of the power-spectrum algorithm. Obtaining the maximum
likelihood power spectrum for the BOOMERanG North America data ---
with ${\cal N}_{p} \sim 2.4 \times 10^{4}$, ${\cal N}_{b} \sim 10$ and
${\cal N}_{l} \sim 1200$ --- requires 50 Gb of disc space, 9.2 Gb of
RAM and $2.8 \times 10^{14}$ flops.

\begin{table*}[ht!]
\centering
\begin{tabular}{cccc}
Calculation & Disc & RAM & Flops \\
\hline
\vspace*{-0.1in}&&&\\
P1 &
$4 \, {\cal N}_{b} \, {\cal N}_{p}^{2}$ &
$16 \,{\cal N}_{p}^{2}$ & 
$O({\cal N}_{l} \, {\cal N}_{p}^{2})$ \\
P2 &
$4 \, ({\cal N}_{b} + 2) \, {\cal N}_{p}^{2}$ &
$16 \,{\cal N}_{p}^{2}$ & 
$\frac{2}{3} \, {\cal N}_{p}^{3}$ \\
P3 &
$4 \, (2 \, {\cal N}_{b} + 1) \, {\cal N}_{p}^{2}$ &
$16 \,{\cal N}_{p}^{2}$ & 
$2 \, {\cal N}_{b} \, {\cal N}_{p}^{3}$ \\
P4 &
$4 \,{\cal N}_{b} \, {\cal N}_{p}^{2}$ & 
$8 \,{\cal N}_{p}^{2}$ & 
$2 \,{\cal N}_{b} \, {\cal N}_{p}^{2}$ \\
P5 &
$4 \, {\cal N}_{b} \, {\cal N}_{p}^{2}$ & 
$16 \, {\cal N}_{p}^{2}$ & 
$3 \, {\cal N}_{b}^{2} \, {\cal N}_{p}^{2}$ \\
P6 &
$4 \, {\cal N}_{b}^{2}$ &
$8 \, {\cal N}_{b}^{2}$ &
$(2 + \frac{2}{3}) \, {\cal N}_{b}^{3}$ \\
\vspace*{-0.1in}&&&\\
\hline
\vspace*{-0.1in}&&&	\\
Total &
$4 \, (2 \, {\cal N}_{b} + 2) \, {\cal N}_{p}^{2}$ &
$16 \, {\cal N}_{p}^{2}$ & 
$(2 \, {\cal N}_{b} + \frac{2}{3}) \, {\cal N}_{p}^{3}$ \\
\end{tabular}
\caption{Computational requirements for each iteration of the power spectrum algorithm}
\end{table*}

Note that in step P1 we calculate the pixel-pixel signal correlation
bin derivative matrices (which we need in step P3 anyway), rather than
the full pixel-pixel signal correlation matrix. Since these are
independent of the bin power step P1 does not need to be repeated at
each iteration. The trade-off here is the near doubling of disc space
required, since we now want to keep these matrices throughout the
calculation, and not overwrite them in step P3. If disc space is at a
premium, however, we can simply revert to recalculation. In step P2 we
calculate all the terms necessary to construct the log-likelihood
itself (\ref{eLL}), which is therefore generated as a useful
by-product, allowing us to confirm that the algorithm is moving to
higher likelihood and to check the appropriateness of our convergence
criterion. In practice over 80\% of the run time is spent in the
triangular solves of step P3. Since these involve level 3 BLAS only
they are highly optimized, and we typically reach 40--80 \% peak
performance on the T3E both at CINECA and at NERSC, with the fraction
slowly decreasing with the total number of processors used

\section{Conclusions}

MADCAP is a highly optimized, portable, parallel implementation (using
ANSI C, MPI and the ScaLAPACK libraries) of the current optimal
general algorithm for extracting the most useful cosmological
information from total-power observations of the CMB. The
$\beta$-release of MADCAP has been ported to a wide range of parallel
platforms --- including the Cray T3E, SGI Origin 2000, HP Exemplar and
IBM SP2. The combination of parallelism and algorithmic optimization
has enabled CMB datasets that would previously have been intractable
to be analyzed in a matter of hours. We have successfully applied
MADCAP at CINECA and NERSC to the data from the North American Test
flight of the BOOMERanG experiment; the scientific results will be
published shortly \cite{B97}.

Existing algorithms are capable of dealing with CMB datasets with up
to $10^{5}$ pixels. Over the next 10 years a range of observations are
expected to produce datasets of $5 \times 10^{5}$ (BOOMERanG LDB),
$10^{6}$ (MAP) and $10^{7}$ (PLANCK) pixels. As shown in Table 3
(where we have assumed a constant 10 multipole bins and 5
Newton-Raphson iterations) the scaling with projected map size pushes
the analysis of these observations well beyond the capacity of even
the most powerful current supercomputers.

\begin{table*}[ht!]
\centering
\begin{tabular}{ccccc}
Flight & ${\cal N}_{p}$ & Disc & RAM & Flops \\ 
\hline 
\vspace*{-0.1in}&&&&\\
\vspace*{0.05in}BOOMERanG NA  &  24,000 & 50 Gb & 9 Gb & $1.4 \times 10^{15}$ \\
\vspace*{0.05in}MAXIMA 1      &  32,000 & 90 Gb & 16 Gb & $3.2 \times 10^{15}$ \\
\vspace*{0.05in}MAXIMA 2      &  80,000 & 560 Gb & 100 Gb & $5.1 \times 10^{16}$ \\
\vspace*{0.05in}BOOMERanG LDB & 450,000 & 18 Tb & 3 Tb & $3.7 \times 10^{18}$ \\
\vspace*{0.05in}PLANCK        & 20,000,000 & 35 Pb & 2.4 Pb & $8 \times 10^{23}$ \\
\end{tabular}
\caption{MADCAP computational requirements for current \& forthcoming CMB observations}
\end{table*}

Such datasets will require new algorithms. The limiting steps in the
above analysis are associated with operations involving the
pixel-pixel correlation matrices for the noise $N$, the signal $S$,
and most particularly their sum $D$. The problem here is the noise and
the signal have different natural bases. The inverse noise
correlations are symmetric, band-diagonal and approximately circulant
in the time domain, while the signal correlations are diagonal in the
spherical harmonic domain. Moreover, thus far we have ignored
foreground sources which will be spatially correlated and so most
naturally expressed in the pixel domain. The fundamental problem is
then that the data correlations --- being the sum of these three terms
--- are complex in all three domains. Developing the approximations
necessary to handle such data sets is an area of ongoing research.

\section{Acknowledgments}

The author would like to thank Amadeo Balbi, Andrew Jaffe, Pedro
Ferriera, Shaul Hanany and Radek Stompor of the COMBAT collaboration,
Xiaoye Li and Osni Marques of the NERSC Scientific Computing Group,
and Paolo de Bernardis, Andrew Lange, Silvia Masi, Phil Mauskopf,
Barth Netterfield, John Ruhl and the other members of the BOOMERanG
team.

\end{document}